\documentclass[conference]{IEEEtran}
\IEEEoverridecommandlockouts
\usepackage{cite}
\usepackage{amsmath,amssymb,amsfonts}
\usepackage{graphicx}
\usepackage{textcomp}
\usepackage{algorithm}
\usepackage{algpseudocode}
\usepackage{booktabs}
\usepackage{siunitx}
\usepackage[table]{xcolor}
\usepackage{colortbl}

\def\BibTeX{{\rm B\kern-.05em{\sc i\kern-.025em b}\kern-.08em
    T\kern-.1667em\lower.7ex\hbox{E}\kern-.125emX}}
    
\begin{document}

\title{A Novel Endorsement Protocol to Secure BFT-Based Consensus in Permissionless Blockchain}
 \author{\IEEEauthorblockN{Ziqiang Xu, Ahmad Salehi S., and Naveen Chilamkurti}
 \IEEEauthorblockA{\textit{Department of Computer Science and Information Technology, La Trobe University, Melbourne, Australia} \\
 Email: \{z.xu, a.salehishahraki, n.chilamkurti\}@latrobe.edu.au}
 }

\maketitle

\begin{abstract}
Permissionless blockchain technology offers numerous potential benefits for decentralised applications, such as security, transparency, and openness. BFT-based consensus mechanisms are widely adopted in the permissioned blockchain to meet the high scalability requirements of the network. Sybil attacks are one of the most potential threats when applying BFT-based consensus mechanisms in permissionless blockchain due to the lack of effective verification mechanisms for participants' identities. This paper presents a novel endorsement-based bootstrapping protocol with a signature algorithm that offers a streamlined, scalable identity endorsement and verification process. This approach effectively safeguards the BFT-based consensus mechanism against Sybil attacks. Using our proposed method, we have conducted thorough security analyses and simulation experiments to assess security, robustness, and scalability advantages in large-scale networks. Our results demonstrate that the scheme can effectively address the identity verification challenges when applying BFT-based consensus in a permissionless blockchain.
\end{abstract}

\begin{IEEEkeywords}
Blockchain, Security, Byzantine Fault Tolerance
\end{IEEEkeywords}

\section{Introduction}

Blockchain technology has found extensive applications across various domains, providing robust support for decentralisation, information security, and privacy protection. Notably, a permissionless blockchain has demonstrated exceptional transparency, openness, and anonymity performance. However, as research on blockchain technology advances, issues related to scalability and sustainability in permissionless blockchain are becoming increasingly apparent. Traditional consensus mechanisms such as Proof of Work (PoW) prioritise security but result in significant resource constraints. Proof of Stake (PoS) is not reliant on computing power but presents challenges for new nodes looking to join midway. Moreover, PoS lacks fairness, and rewards tend to favour users with majority stakes. 

Byzantine Fault Tolerance (BFT) based consensus is achieved by multiple nodes communicating and coordinating with each other to reach a consensus decision. This algorithm ensures that the security and consistency of the network are maintained even when there are some faulty or malicious nodes in the network. According to the survey in~\cite{yadav2023comparative}, BFT-based consensus mechanisms have lower computational device performance demands and operational costs compared to PoW. Therefore, they offer good scalability and are suitable for lightweight networks. Previous literature has attempted to apply BFT-based consensus in permissionless blockchains like Eyal et al.~\cite{eyal2016bitcoin} introduced Bitcoin-NG, a Byzantine Fault Tolerant blockchain protocol, Kokoris-Kogias et al.~\cite{kogias2016enhancing} applied Practical Byzantine Fault Tolerance (PBFT) to develop ByzCoin, a cryptocurrency. These schemes experimentally demonstrate the performance (throughput, latency) and scalability advantage of a BFT-based consensus mechanism applied in a permissionless blockchain. However, the literature~\cite{kogias2016enhancing} mentions the challenge of preventing Sybil attacks during their implementation. Sybil attacks involve malicious users creating multiple false identities to gain control over a majority of nodes or resources, disrupting normal network operations or manipulating transaction processes. In a permissioned blockchain, the identity can be verified by a trusted third party such as Certificate Authority (CA) to prevent false identities in the network \cite{alhajri2022blockchain, salehi2023access}. In a permissionless blockchain, other Sybil attack-resistant methods are required to guarantee that the nodes in the network are trustworthy without destroying the openness. 

One approach to resist Sybil attacks is to utilise a PoW and its variants as a pre-consensus mechanism for bootstrapping. The PoW raises the cost of implementation of Sybil attacks through extensive computation. Several schemes, such as Elastico~\cite{luu2016secure}, Omniledger~\cite{kokoris2018omniledger}, and RapidChain~\cite{zamani2018rapidchain}, have developed hybrid PoW systems that effectively defend against Sybil attacks by setting the appropriate difficulty for the PoW algorithm. A study by ~\cite{wang2022bft} suggests that PoW difficulty level settings are typically less competitive than in Bitcoin. This is because setting the difficulty of PoW requires balancing fairness between different performance devices, and also guarantees the operational cost and bootstrapping efficiency in various application scenarios. Another approach utilises PoS and its variants, such as Poster~\cite{lee2019poster}, which designs PoS-based methods for the equitable selection of consensus nodes. Ouroboros~\cite{kiayias2017ouroboros} proposes a method that periodically assigns block producers at random. Algorand~\cite{gilad2017algorand} assigns weights to each user based on the number of tokens in their wallet through public randomness. These PoS methods increase the cost of creating false identities through stakes and provide a lower operating cost than PoW. However, the scalability and efficiency of these methods in the bootstrapping process have not been fully tested, and the~\cite{gilad2017algorand} method currently only applies to cryptocurrencies and not general-purpose blockchains. Therefore, achieving an efficient, secure, and scalable bootstrapping process in permissionless blockchain remains a big challenge.

This paper addresses the challenges of securely implementing a BFT-based consensus mechanism in a permissionless blockchain. We set up an endorsement group for each participating node and endorse their deposit  (like monetary value or stocks) to prevent Sybil attacks. This protocol uses an aggregated signature scheme based on Boneh–Lynn–Shacham (BLS)~\cite{boneh2001short} to achieve a secure and efficient endorsement verification. Meanwhile, the efficient endorsement group scheme reduces the cost and time of the bootstrapping process.

Our scheme offers significant advantages in the following aspects. \textbf{Resource Efficiency:} Ensuring node security defends against Sybil attacks and substantially reduces resource consumption during the endorsement process. \textbf{Adaptability and Scalability:} The communication complexity does not exponentially grow regardless of network size, and it works efficiently even on light computational devices such as IoTs. \textbf{Openness and Fairness:} No third party is required for authentication. Any node that meets the minimum conditions can join at any time to participate in the consensus.

The rest of the paper is organised as follows. We introduce the model and preview the scheme in Section \ref{Scheme Overview}. Section \ref{Node Bootstrapping}, describes the endorsement group selection and bootstrapping process in detail. Section \ref{Endorsement Protocol} describes the endorsement protocol and proves the protocol's security. Section \ref{Endorsement Group Size Security} performs a security analysis of the scheme. Section \ref{Performance Evaluation} evaluates scalability and performance. Section \ref{Conclusion} summarises and gives conclusions.

\section{Model and Overview}
\label{Scheme Overview}

\subsection{Scheme Overview}
Our scheme validates the key pair $(sk, pk)$ by verifying the authenticity of the node deposit and signing it with endorsement. The key pair is the node's identity; only a valid key pair can participate in consensus. The scheme has the following key components:
\begin{itemize}
  \item \textbf{Global State Ledger:} All participating nodes maintain a distributed ledger that is the global state ledger known as the main blockchain. This ledger records node deposits and all valid endorsement signature information and ensures the node keys' validity.

  \item \textbf{Endorsement Groups:} The endorsement group is designed to verify the deposit's authenticity and validate or invalidate the key pair $(sk, pk)$. We use a BFT-like principle for the endorsement process. When nodes get over 2/3 of endorsing nodes' approval, they can activate their key pair $(sk, pk)$.

  \item \textbf{Verifiable Aggregate Signatures:} A BLS-based aggregate signature mechanism that reduces the complexity of signature verification and provides significant advantages for node authentication, fast transaction processing, and overall system performance.
  
\end{itemize}

\subsection{Network Model}

In our scheme, communication within the network relies on a partially synchronous gossip protocol~\cite{karp2000randomized}. This protocol guarantees that messages sent by honest nodes will reach all other honest nodes within a fixed time interval $\Delta$, although it may not preserve the order of these messages. Regarding the endorsement protocol, we assume that communication connections between nodes and their endorsing nodes are reliable. This means that any node can obtain the returned information within a maximum time of $2\Delta + P$, where $P$ is the maximum execution time of the protocol.

\subsection{Threat Model}
Our threats model consists of Sybil attacks, which will influence network fault tolerance. Sybil attacks involve malicious users using various means to create multiple false identities. Network fault tolerance is defined as assuming that there are a number of f fault nodes in the network, then the number of participating nodes in the network is at least not less than 3f+1. That fault node includes (1) Node disconnection and network interference, where nodes may disconnect from the network at any time due to internal faults or network issues, and (2) Adversarial attacks, where adversaries send invalid or inconsistent messages during any consensus phase or choose to remain silent in an attempt to disrupt the consensus process.

\section{Node Bootstrapping} 
\label{Node Bootstrapping}
Node bootstrapping introduces the process of nodes joining the network. This includes selecting an endorsement group for each node and performing the endorsement.

\subsection{Endorsement Group Selection}
\label{Sharded Endorsement Group}
We introduce the concept of endorsement groups to achieve secure node bootstrapping. This approach randomly allocates a dedicated endorsement group for each potential consensus node, denoted as $N_c$. For this purpose, we have devised a two-step random selection methodology, significantly reducing the probability of endorsement group failure due to its parallel duplicate selections.

\begin{figure}[ht]
    \centering
    \includegraphics[width=9cm]{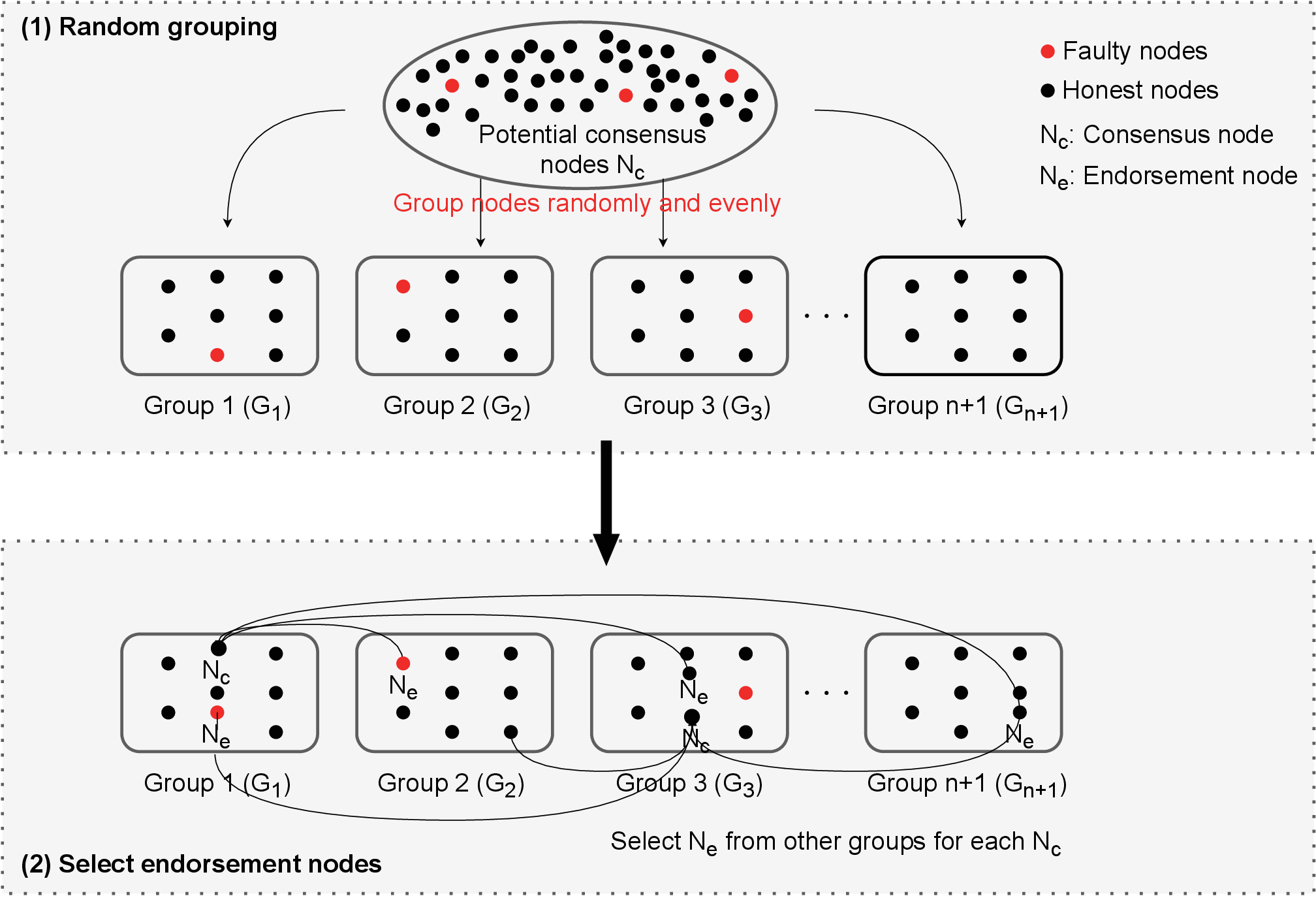}
    \caption{Select Endorsement Node}
    \label{fig: Node Selection}
\end{figure} 

Fig. \ref{fig: Node Selection} illustrates the specific selection approach. The set of $N_c$ entities within the network are uniformly shuffled and evenly divided into $n+1$ candidate subgroups, denoted as $G_i$ (where $i\in(1, n+1)$). Each subgroup contains $k$ instances of $N_c$. According to our threat model, we assume that the number of corrupt nodes (denoted as $N_{cF}$) satisfies $N_{cF} < \frac{1}{3}N_c$, and the number of honest nodes (denoted as $N_{cH}$) satisfies $N_cH \geq \frac{2}{3}k$. Given that this grouping process conforms to a hypergeometric distribution, approximately $80\%$~\cite{zamani2018rapidchain} of the groups will have a proportion of fault nodes lower than one-third. For each node $N_c$ in group $G_i$, we randomly select one node from all the other groups as its endorsement node. Therefore, each $N_c$ will be assigned $n$ endorsement nodes $N_e$. When randomly selecting $n$ nodes, we denote the probability that among the $n$ nodes chosen, $N_{cF} > \frac{1}{3}n$ of selection failure as $Pr[F]$. The value of $N_{cF} > \frac{1}{3}n$ with $n$ choices can be expressed as \eqref{eq1}.
\begin{equation} \label{eq1}
  \sum^{\frac{2n}{3}}_{i=1} (\frac{k}{2})^{\frac{n}{3} + i}\times(\frac{k}{2})^{\frac{2n}{3} - i}
\end{equation}
The probability of failure $Pr[F]$ can be expressed as \eqref{eq2}.
\begin{equation} \label{eq2}
  Pr[F] = \frac{\sum^{\frac{2n}{3}}_{i=1} (\frac{k}{2})^{\frac{n}{3} + i}\times(\frac{k}{2})^{\frac{2n}{3} - i}}{k^n}
\end{equation}

In \eqref{eq1} and \eqref{eq2}, we assume that the percentage of fault nodes in each group of $k$ nodes is maximum $\frac{1}{3}$. By our calculations in \ref{Endorsement Group Size Security}, even if the percentage of fault nodes reaches $\frac{1}{2}$, a suitably chosen value for n (more than 20 nodes) guarantees an extremely low failure rate $Pr[F]$. Furthermore, we use Epoch and reconfiguration concepts similar to \cite{zamani2018rapidchain} to deal with nodes joining in the middle of the epoch. When a node joins in the middle of $Epoch_(n-1)$, it randomly selects $n$ groups from the $n+1$ candidate groups and assigns them to the endorsement group. Subsequently, the node is randomly assigned to the $n+1$ candidate group at the next $Epoch_(n)$. This method ensures that adding new nodes does not change the failure probability of the candidate group. Therefore, under normal network conditions (absent sudden bursts of node influx), the workload during the reconfiguration phase between Epochs will be significantly lower than that of the initial bootstrap phase.

\subsection{Endorsement Processes}
The endorsement process will begin after selecting the endorsement group for each potential participating node. During the endorsement process, each $N_c$ will submit a participation request $\text{Tx:} \{ \text{performance matrix}, \text{token address} \}$ to its endorsement nodes $N_e$, where the performance matrix evaluates whether or not $N_c$'s performance can run the consensus mechanism. The token address is the participant's individual deposit used to endorse their key pairs. The deposit amount depends on the node's potential benefits, and to defend the Sybil attack, the deposit for $N_c$ typically exceeds the benefits obtained in a few rounds. When $N_e$ receives the $Tx$, they will verify the token's authenticity and evaluate the performance matrix. For compliant $N_c$, $N_e$ makes an endorsement signature of $N_c$'s token address and public key and returns it to $N_c$. When the number of endorsement signatures collected by $N_c$ exceeds $\frac{2n}{3} +1$, it becomes a consensus node and publishes its signature to the main blockchain chain. 

In the endorsement process, each node communicates with $n$ other nodes, exchanging a fixed number of requests and responses in parallel, so the communication complexity is $O(n)$ and the storage complexity is $O(\frac{|b|}{n})$. At any given time, when a node initiates a legitimate exit request, the network will reassign the endorsing node and repeat the verification of the endorsement process to invalidate the node's key pairs. Furthermore, we establish the constraint that nodes with identical identities can only submit an endorsement request once per Epoch, thereby ensuring protection against malevolent node attacks on the network.

\subsection{Security Settings and Arguments}
The selection of endorsement nodes for each $N_c$ can be achieved through the bias-resistant distributed randomness as demonstrated in Omniledger~\cite{kokoris2018omniledger}. This approach helps mitigate the risks associated with insecure behaviour resulting from nodes autonomously selecting their preferred endorsement nodes. Furthermore, similar to Byzantine Fault Tolerance (BFT), we have provided evidence in Sec.~\ref{Endorsement Group Size Security} that there is an extremely low probability of failure in node selection for any honest node. Therefore, it can be considered a fair selection process that does not decrease the proportion of honest nodes due to the selection mechanism. 

In this work to effectively use deposits to resist the Sybil attack, we must ensure that the deposits are authentic, verifiable and non-reusable. The literature has shown that their research work done in this area. For example, we can use the InterPlanetary File System (IPFS)~\cite{benet2014ipfs} scheme to store the deposit, which cryptographic properties can secure. Secondly, we prove in Sec.\ref{Endorsement Protocol} that the endorsement signature process is secured. Thus, we can consider this scheme to be an effective Sybil attack-resistant scheme.

\section{Endorsement Protocol}
\label{Endorsement Protocol}

We design BLS-based aggregated signature schemes for our endorsement protocols. The inspiration for this scheme draws from the multi-signature scheme (MSP) and aggregation multi-signature scheme (AMSP) proposed by Boneh~\cite{boneh2018compact}, utilising signature aggregation to reduce the signature size and verification complexity. The operational principles of our specific signature scheme are as follows:

\textbf{Preliminaries.} Assuming a non-degenerate, efficiently computable, bilinear pairing $e: \mathbb{G}_1 \times \mathbb{G}_2 \rightarrow \mathbb{G}_t$, where $\mathbb{G}_1$, $\mathbb{G}_2$, $\mathbb{G}_t$ are groups of prime order $q$. Let $g_1$ and $g_2$ be generators of $\mathbb{G}_1$ and $\mathbb{G}_2$ respectively \cite{shahraki2020attribute}. 

\textbf{Definition 1.} Hash functions: $H_0 : \{0, 1\}^* \rightarrow \mathbb{G}_1$, and $H_1 : \{0, 1\}^* \rightarrow \mathbb{Z}_q$.

\textbf{Definition 2.} View $V_i: f(pk_i, \mathbb{PK})$ represents a view associated with a specific public key $pk_i$, where $\mathbb{PK} = \{pk_1, pk_2, \ldots, pk_n\}$ constitutes a set of public keys.

\textbf{Definition 3.} Vector $c_i: [pk_1=0, pk_2=1, \ldots, pk_n=0]$, where the set of $pk$ are all elements of $\mathbb{PK}$. Each entry of vector $c_i$ is assigned by the mapping function $f: \{\text{true}, \text{false}\} \rightarrow \{1, 0\}$, wherein true is mapped to 1, and false is mapped to 0.

\textbf{Key Generation.} The key generator algorithm, $KeyGenerator()$, generates $sk$ and $pk$ through the employment of \eqref{eq5}.
\begin{equation}\label{eq5}
\begin{aligned}
sk & \leftarrow \text{random}(\mathbb{Z}_q) \
pk & \leftarrow g_2^{sk} \in \mathbb{G}_2
\end{aligned}
\end{equation}

\textbf{Signing.} The signature algorithm, $Sign(sk_i, m, PK)$, produces the signature $s_i$ during the signing process, with its computation outlined as \eqref{eq7}.
\begin{equation}\label{eq7}
    s_i \leftarrow H_0(m)^{p_i \cdot sk_i}
\end{equation}
Where $p_i \leftarrow H_1(pk_i, \mathbb{PK})$, $m$ represents the message signed, while $sk_i$ and $pk_i$ denote the signer's key pair.

\textbf{Signature Aggregation.} The aggregated signature, $SignAgg(c_i, \{s_1, s_2, \ldots, s_n\})$, combines all collected signatures $s_i$ and produces the aggregated signature $\sigma$.
\begin{equation}\label{eq8}
\sigma \leftarrow \prod_{j \in y} s_j
\end{equation}

where $y = c_i: \{pk_i | pk_i = 1\}$ denotes the set of all $pk$ in $c_i$ with $pk_i = 1$.

\textbf{Public Key Aggregation.} Aggregating public keys is a critical component for verifying the authenticity of signatures. $PKAgg(c_i, \mathbb{PK})$ generates the aggregated public key $a$.
\begin{equation}\label{eq9}
    a \leftarrow \prod_{j \in y} pk_j^{H_1(pk_j, \mathbb{PK})}
\end{equation}

\textbf{Verification.} $Verify(\sigma, a, m)$ outputs success when Equation \eqref{eq10} holds.

\begin{equation}\label{eq10}
    e(\sigma, g_2) = e(H_0(m), a)
\end{equation}

With the signature scheme, Algorithm \ref{alg:alg1} presents the pseudocode for the endorsement protocol.

\begin{algorithm}
\caption{Endorsement Protocol}\label{alg:alg1}
\begin{algorithmic}
\State \textbf{Key Generation:}
\State (1) $N_c$ invokes $KeyGenerator()$ to generate $(sk, pk)$.
\State \textbf{Endorsement Preparation:}
\State (1) Each node endorsement group view $V_i: f(pk_i, \mathbb{PK})$ is generated by the selection step of Sec III-A. Where $\mathbb{PK}$ is the set of public keys for the selected endorsement nodes.
\State (2) $N_c$ converts the $V_i$ into an endorser count vector $c_i$.
\State (3) $N_c$ sends an endorsement request $Tx:$ \{performance matrix, token address\} to $N_e$.
\State \textbf{Endorsement Signing:}
\State (1) $N_e$ verifies $Tx$ and generates a signature message $m:\{pk_i, token address\}$. \State (2) Then, $N_e$ employs $Sign(sk_i, m, \mathbb{PK})$ to generate the signature $s_i$, where $sk_i$ is $N_e$'s private key and $\mathbb{PK}$ is $N_c$'s public key set.
\State \textbf{Signature Aggregation:}
\State (1) When $N_c$ collects a signature from $N_e$, it marks 1 in the corresponding $c_i$.
\State (2) When count($c_i=1$) $> \frac{2}{3}$count($c_i$), $N_c$ invokes $SignAgg(c_i, \{s_1, s_2, \ldots, s_n\})$ to generate the aggregated signature $\sigma$ and announces $c_i$ to the network.
\State \textbf{Signature Verification:}
\State (1) Any node in the network can verify $\sigma$: (a) $PKAgg(c_i, \mathbb{PK})$ generates the aggregated public key $a$, where $c_i$ and $\mathbb{PK}$ are publicly available. (b) Verify by using $Verify(\sigma, a, m)$.
\end{algorithmic}
\end{algorithm}

\textbf{Proof of Signature Security:} Regarding the security of signatures, we will establish their validity from two perspectives. (1) Signatures $\sigma$ aggregated by honest nodes are verifiable by any node. (2) Malicious nodes are incapable of forging signatures of honest endorsers.

(1) Given that $\mathbb{G}_1$, $\mathbb{G}_2$, $\mathbb{G}_t$ are cyclic groups belonging to the commutative group. It is possible to prove that the equation can be verified knowing $\sigma$, $\mathbb{PK}$, and $c_i$ by the following equation:

\begin{align*}
    e(\sigma, g_2) &= e\left(\prod_{j \in y} s_j, g_2\right) = e\left(\prod_{j \in y} H_0(m)^{sk_j \cdot p_j}, g_2\right) \\ 
    &= \prod_{j \in y} e\left(H_0(m)^{sk_j \cdot H_1(pk_j,\mathbb{PK})}, g_2\right) \\
    &= \prod_{j \in y} e\left(H_0(m), g_2^{sk_j \cdot H_1(pk_j,\mathbb{PK})}\right) \\
    &= \prod_{j \in y} e\left(H_0(m), pk_j^{H_1(pk_j,\mathbb{PK})}\right) \\
    &= e\left(H_0(m), \prod_{j \in y} pk_j^{H_1(pk_j,\mathbb{PK})}\right) \\
    &= e\left(H_0(m), a\right)
\end{align*}

(2) In the proposed threat model, adversarial nodes possess two means of forgery: (a) Forging signatures of f+1 honest nodes. (b) Rogue Public-Key attack.

(a) For signature forgery, we consider a rare scenario where the total number of nodes is n=3f+1, and f adversary nodes exist. In this scenario, a number of f adversary nodes collude to assist one of them in forging the signatures of the remaining honest nodes. Consequently, the f adversaries must forge f+1 signatures, with f-1 nodes contributing to forging one signature each and another forging two signatures. Let us begin by discussing the case of forging a signature for a single honest node.

Suppose we aim to forge the signature of $sk_1$, requiring us to find an $s_1$ such that $e(s_1,g_2) = e(H_0(m),pk_1^{H_1(pk_1,\mathbb{PK})})$. Where $pk_1 = g_2^{sk_1H_1(pk_1,\mathbb{PK})}$, rearranging the equation, we arrive at $s_1 = H_0(m)^{sk_1H_1(pk_1,\mathbb{PK})}$. When $H_1(pk_1,\mathbb{PK})$ is known, forging $s_1$ necessitates finding an integer $z$ that satisfies $z \cdot H_1(pk_1,\mathbb{PK}) \equiv 1 \pmod{q}$. However, discovering an integer $z$ to ensure that $s_1 \in G_1$ and satisfies $e(s_1,g_2) = e(H_0(m),pk_1^{H_1(pk_1,\mathbb{PK})})$ is the Computational Diffie-Hellman (CDH) problem. Consequently, accomplishing this task within a given time is virtually infeasible.

(b) In the context of rogue attacks, when colluding with $f$ nodes, in order to achieve equation \eqref{eq10}, it is necessary to possess a rogue public key $pk \in \mathbb{G}_2$ to make the equation \eqref{eq11} holds.
\begin{equation}\label{eq11}
    pk^{H_1(pk, \mathbb{PK})} = g_2^{\alpha} \cdot \prod_{j \in y} pk_j^{-H(pk_j, \mathbb{PK})}
\end{equation}
where $\alpha \leftarrow \mathbb{Z}_q$ is chosen by the adversary. Then declaring $\sigma = H_0(m)^{\alpha}$ and $pk$ which can makes the equation \eqref{eq10} holds, because 

\begin{align*}
    e(\sigma, g_2) &= e(H_0(m)^{\alpha}, g_2) = e(H_0(m), g_2^{\alpha}) \\
    &= e(H_0(m), \prod_{i \in y} pk_i^{H(pk_i,\mathbb{PK})}) = e(H_0(m), a)
\end{align*}

However, solving for $pk$ in equation \eqref{eq11} involves a discrete logarithmic problem that is difficult to solve in a reasonable time. 

\section{Endorsement Group Size}
\label{Endorsement Group Size Security}

In our endorsement group, we can reuse nodes for endorsing different $N_c$, which means the endorsement group can be small. This means the proportion of malicious nodes for each candidate group does not change during bootstrapping. This way, the probabilities calculated according to equation \eqref{eq2} are not biased.

In Table \ref{tab:Failure Probability for Different Numbers of Candidate Groups}, we compute the endorsing group's probability of failure for different numbers of endorsement nodes with $1/2$ and $1/3$ adversary ratios. The endorsement group failure means that the number of malicious nodes in the group is greater than $1/3$.

\begin{table}[ht]
\centering
\caption{Failure Probability for Different Numbers of Endorsement Nodes}
\label{tab:Failure Probability for Different Numbers of Candidate Groups}
\begin{tabular}{>{\columncolor[gray]{0.9}}c S[table-format=1.4e-2] S[table-format=1.4e-2]}
\toprule
\rowcolor[gray]{0.9} Number of groups & {$N_{cF} = 1/2k$} & {$N_{cF} = 1/3k$} \\
\midrule
10 & \num{5.8594e-03} & \num{1.6936e-03} \\
11 & \num{3.4180e-03} & \num{9.0326e-04} \\
12 & \num{1.9531e-03} & \num{4.7983e-04} \\
13 & \num{9.7656e-04} & \num{2.5389e-04} \\
14 & \num{5.4932e-04} & \num{1.3461e-04} \\
15 & \num{3.0518e-04} & \num{7.1295e-05} \\
16 & \num{1.5259e-04} & \num{3.7724e-05} \\
17 & \num{8.3923e-05} & \num{1.9971e-05} \\
18 & \num{4.5776e-05} & \num{1.0570e-05} \\
19 & \num{2.2888e-05} & \num{5.5929e-06} \\
20 & \num{1.2398e-05} & \num{2.9598e-06} \\
21 & \num{6.6757e-06} & \num{1.5662e-06} \\
22 & \num{3.3379e-06} & \num{8.2873e-07} \\
23 & \num{1.7881e-06} & \num{4.3852e-07} \\
24 & \num{9.5367e-07} & \num{2.3204e-07} \\
25 & \num{4.7684e-07} & \num{1.2278e-07} \\
26 & \num{2.5332e-07} & \cellcolor{red!25}\num{6.4968e-08} \\
27 & \num{1.3411e-07} & \num{3.4377e-08} \\
28 & \cellcolor{red!25}\num{6.7055e-08} & \num{1.8190e-08} \\
29 & \num{3.5390e-08} & \num{9.6249e-09} \\
30 & \num{1.8626e-08} & \num{5.0929e-09} \\
\bottomrule
\end{tabular}
\end{table}

According to the calculation results, we can see that the probability of failure decreases as the number of endorsement nodes increases. The data marked in red shows that for $\frac{1}{2}$ and $\frac{1}{3}$ of the adversary ratio, the failure probability will be less than $10^{-7}$ when the number of endorsement nodes is greater than 28 and 26 groups, respectively. We utilise the probability density function of the exponential distribution: $f(x) = \lambda e^{-\lambda x}$, where $\lambda$ represents the event occurrence rate. The average expectation $E[X] = \frac{1}{\lambda}$ signifies the average times to failure. This implies that if the failure probability is less than $10^{-7}$, for every allocation of ten million endorsement groups, there is a possibility of encountering only one failed group.

Based on this result, we conclude that when the nodes are evenly divided into more than 28 groups under a restricted threat model, the adversary nodes do not pose damage to the endorsement group. It is also worth noting that the number of nodes in an endorsement group should be kept to a minimum number that does not pose a threat to reduce the endorsement pressure on each endorsing node and thus shorten the time spent on the endorsement process.

\section{Performance Evaluation}
\label{Performance Evaluation}
\subsection{Bootstrapping Performance}
We simulated a virtual environment with constrained network conditions using Python to evaluate the performance of our proposed bootstrapping process. To simulate the real environment, we limited the communication rate between nodes to no more than 25 Mbps and introduced a delay of 100 ms for the network links. In our simulations, we reproduce the bootstrapping process, which starts with a sufficient number of nodes in the network and then performs endorsement node selection and the final endorsement process. We simulated using 2000, 4000, 8000, and 10000 nodes and randomly introduced a certain number of malicious nodes among them.

\begin{figure}[ht]
    \centering
        \includegraphics[width=9.85cm]{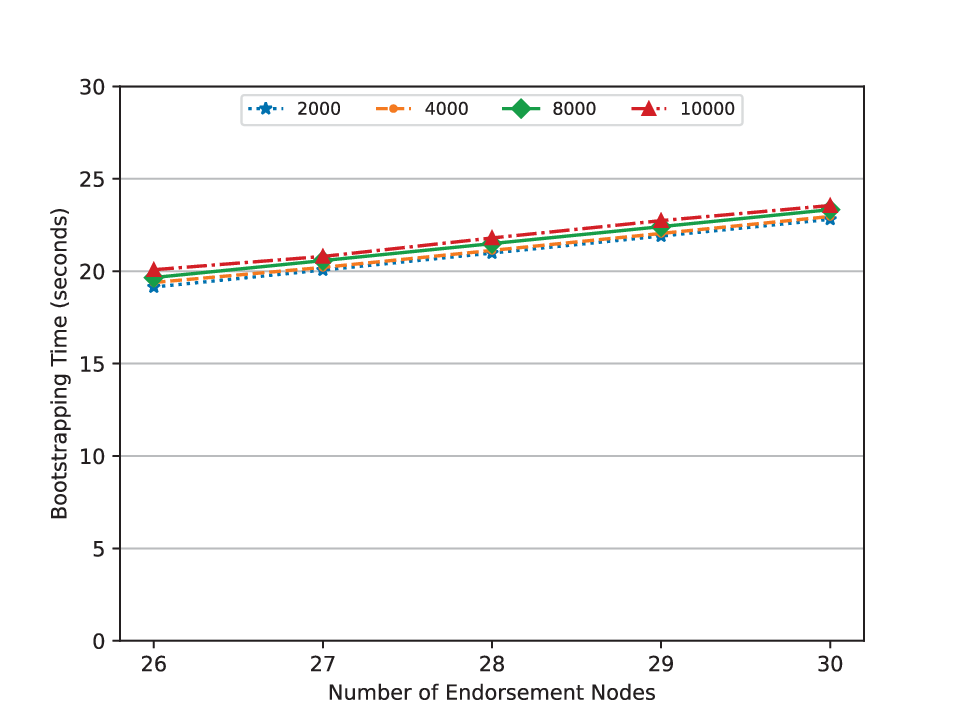}

    \caption{Time required to complete bootstrapping for different numbers of total nodes with different numbers of endorsement nodes set up}
    \label{fig: Boostrapping}
\end{figure} 

Fig. \ref{fig: Boostrapping} shows the maximum time required for all nodes in the network to complete the bootstrapping process with a different number of endorsement nodes. It is the time from the operation of the whole network to the completion of the endorsement by the last node. We compared our results with RapidChain~\cite{zamani2018rapidchain}. Since RapidChain does not evaluate the bootstrapping process separately, we focus on the experimental results of its reconfiguration process. Their experimental results contain three components: reference committee generation of epoch randomness, consensus of proposed new configuration blocks and assignment of new nodes to existing committees. The first two of those parts are the process of bootstrapping nodes using PoW consensus. Based on the experimental results, it can be seen that the bootstrapping time of RapidChain with 4000 nodes in the network is about 75.2 seconds. In comparison, the bootstrapping time of our scheme is between 19.38 seconds and 22.96 seconds. Meanwhile, according to our experimental results, it is shown that in our scheme, as the number of endorsement nodes increases, the time taken by the nodes to complete the bootstrapping process also increases. This is because, with more endorsement nodes, $N_c$ needs to communicate with more endorsing nodes. However, despite the increase in the number of endorsement nodes, the increase in bootstrap time is relatively slow, indicating the scalability of our approach in dealing with large-scale networks. On the other hand, we observe that the bootstrap time increases accordingly as the number of nodes increases. However, the growth of bootstrap time shows a linear growth trend rather than exponential growth. This suggests that our method can maintain a consistent bootstrapping performance in the face of large-scale node networks, which is crucial in real-world applications.

\subsection{CPU Utilisation}
To deeply explore the adaptability of our scheme to lightweight users and its effect on energy savings, we compare CPU utilisation when using the PoW mechanism and our proposed approach. The purpose of these experiments is to verify whether our approach is effective in reducing the burden on computing resources in large networks. We fixed the number of CPU compute cores of the nodes in our experiments and optimised the computation process to ensure PoW and our proposed scheme are completed in the shortest possible time. We monitored the CPU utilisation while running PoW and our methods and presented the results in the following two graphs.

\begin{figure}[ht]
    \centering
    \includegraphics[width=9.4cm]{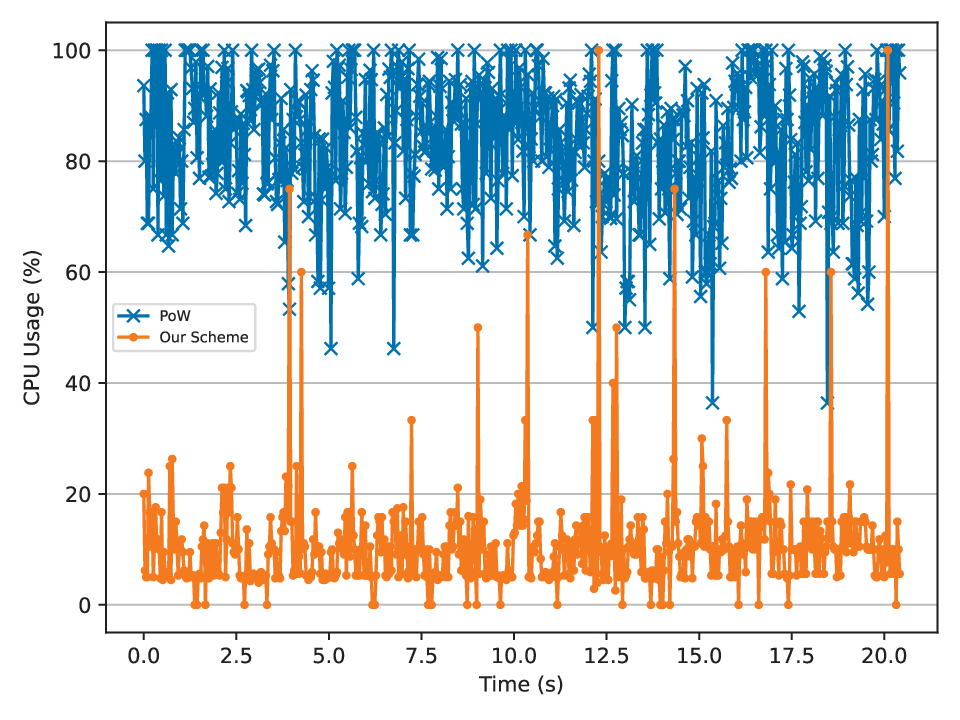}
    \caption{CPU utilization of PoW}
    \label{fig: cpu_usage_p2}
\end{figure} 

Fig. \ref{fig: cpu_usage_p2} shows the CPU utilisation data captured during the experiments for 20 seconds. It is important to note that PoW may take longer in real-world situations due to different difficulty settings, so we have adapted the computational process in our experiments to be completed within 20 seconds. Based on the results of the graphs, it is obvious that our method shows a significant advantage in terms of CPU utilisation. Compared to the PoW mechanism, our method can significantly reduce the CPU workload. This means that our approach requires less device performance when deploying a blockchain network, enabling better engagement of lightweight devices and reducing energy consumption. This further demonstrates the superior performance of our scheme in improving system scalability and resource utilisation efficiency.

\section{Application Examples}
\label{Application}
This section discusses three application scenarios where our approach is applicable. We illustrate in detail the potential of this endorsement protocol during its practical application.

\subsection{Sharding Blockchain}

Sharding techniques have been widely used in permissionless blockchain systems to improve scalability and throughput. This is because it involves selecting nodes with the required capabilities to maintain the consistency of the sharding and to secure the network. Traditional solutions, such as Proof of Work (PoW), Proof of Stake (PoS), and Proof of Burning (PoB), while effective in guarding against malicious behaviour, are often complex to build enough trust and set the right incentives to ensure proper node functioning as well as network security. However, it is difficult for these solutions to build enough trust and develop appropriate incentives to ensure the adequate functioning of nodes. Our approach can be critical in the sharding node selection phase, especially in preventing Sybil attacks. Nodes qualify as verifiers of the sharding network by depositing money rather than wasting computational resources. The endorsement protocol ensures that the elected nodes have sufficient computational power and resources to maintain the shard blockchain and that their trusted identities are readily verifiable. For example, it can replace the mechanism of resisting Sybil attacks of Omniledger~\cite{kokoris2018omniledger} and RapidChain~\cite{zamani2018rapidchain} and realize the guarantee that the network structure is not changed in bootstrapping nodes. It reduces the bootstrap time and enhances the security and reliability of the whole system and the integrity of the blockchain. This helps to provide scalability and trustworthiness for the broader blockchain ecosystem and a better experience for users and developers.

\subsection{Permissionless Blockchain}

Scalable permissionless blockchain systems usually face the challenge of adopting lightweight consensus mechanisms based on Byzantine Fault Tolerance (BFT). This is because authentication between nodes is challenging for permissionless environments. Our approach provides strong support for applying BFT-based consensus mechanisms in permissionless blockchains. By performing pseudo-authentication, we can ensure that only legitimate nodes are eligible to participate in the consensus process. Endorsement protocol significantly reduces the impact of malicious nodes and implements a lightweight consensus mechanism based on the Byzantine Fault Tolerance (BFT) model. In public blockchains such as ByzCoin~\cite{kogias2016enhancing}, for example, by integrating an endorsement protocol, these networks can efficiently verify node identities, thus enhancing the security and reliability of their consensus protocols. This innovation opens the door to many critical use cases, such as industrial~\cite{xu2022blockchain} and decentralized applications~\cite{xu2023blockchain}.

\subsection{Permissioned Blockchain}

The secure signature design of the endorsement protocol is also applicable in permissioned blockchains that require large-scale verification. The algorithm can efficiently help Certification Authorities (CA) verify trusted identity. For example, when different organization administrators or CAs exchange identity certificates in a consortium blockchain, the authenticity of the verification can be verified more efficiently by the aggregated signature verification algorithm. In addition, our approach provides better scalability as it can automatically perform authentication in a blockchain network without the intervention of a centralized authority. This offers more possibilities for various application scenarios such as authentication, digital identity management and access control \cite{salehi2023dacp}.

\section{Conclusion}
\label{Conclusion}

This paper presents a novel solution for applying a secure BFT-based consensus mechanism in permissionless blockchains. The proposed scheme provides a secure, fair and efficient defence against Sybil attacks. We establish the proposed protocol's effectiveness and scalability through detailed design and analysis. Our performance evaluation results demonstrate that, in large-scale networks, the protocol can reduce CPU utilisation and enhance system efficiency. Overall, this research overcomes the scalability and sustainability issues when bootstrapping using PoW, provides an innovative approach for the widespread use of permissionless blockchains, and makes significant progress regarding security, efficiency, and scalability.

\end{document}